# UXR Point of View on Product Feature Prioritization Prior To Multi-Million Engineering Commitments

Jonas Lau
Google LLC
Mountain View
jonaslau@google.com

Annie Tran
Google LLC
Mountain View
anntran@google.com

## ABSTRACT

This paper discusses a popular UX research activity, feature prioritization, using the User Experience Research Point of View (UXR PoV) Playbook framework. We describe an application of multinomial logistic regression, frequently marketed as MaxDiff, for prioritizing product features in consumer product development. It addresses challenges of traditional surveying techniques. We propose a solution using MaxDiff to generate a reliable preference list with a reasonable sample size. We also adapt the MaxDiff method to reduce the number of survey responses in half, making it less tedious from the survey takers' perspective. We present a case study using the adapted MaxDiff method for tablet feature prioritization research involving users with disabilities.

## INTRODUCTION

When a technology company decides to launch a new or an updated product, the product management (PM) team would usually set up some programs to establish a business case for the new products (e.g., Tetlock & Gardner, 2015). In these programs, different aspects of the current and new products would be assessed. For the product to succeed, the new product has to be competitive in terms of pricing, feature offerings, and technological advancements.

User experience (UX) teams, including UX design and UX research teams, would be involved in brainstorming new product features. These new features would hopefully address customer needs and bring a new competitive product to the market. This paper focuses on this early feature selection stage of the product development cycle. Later stages of the cycle, such as feature development, implementation, and testing, are outside the scope of the current discussion.

To better structure our proposal, we employ the User Experience Research Point of View (UXR PoV) Playbook to guide us through the development of a "play" (Dogan, Giff, & Barsoum, 2024). Specifically, we will present our case study as a play, while elaborating on the issue, type, related cards, and best practice of the play.

## Resource Constraints in Product Development

In consumer product development, there are usually multiple brainstorming sessions for product management teams to solicit new product feature ideas (e.g., components and features for a smart home product). This is the same regardless of whether the product is hardware- or software-based (e.g., a security camera in the former, and a payment app in the latter case).

A laundry list of potential new features is likely a result of these brainstorming sessions. It is almost certain that the company cannot afford to invest in all the new features proposed in the brainstorming sessions due to the high time- and engineering resource requirements. A user experience researcher (UXR) is then brought into the team to test each feature and see which resonates more with the users. This is done when other teams are assessing the financial or technological feasibility of the proposed features.

## Mental Constraints on Research Participants

When doing research on potential product features, one popular approach is "contextual inquiry". Research participants are asked to perform some tasks in their natural environment, with the UX researchers observing the behaviors (Beyer & Holtzblatt, 1990). As part of the study, the researchers may present the list of potential product features to the participants, and ask for feedback on each feature. With a list of 20 features, participants are expected to either rate the usefulness of each feature, or rank the features in terms of their perceived usefulness.

These evaluation tasks can be very mentally taxing for the participants, given the time constraints to evaluate these features in a UX study session. An important implicit assumption of the evaluation task is that participants have the mental capacity to hold the information concurrently. They can compare and rate each feature in comparison to the others. With recent decades of behavioral economics research, this assumption seems to have its limitations. According to the research on bounded rationality (Simon, 1982), these UX participants are likely giving the researchers a satisfactory solution, rather than an optimal solution.

At the end, in each new generation of a product, only a handful of new features ultimately make it into production. These new features are likely to be promoted in huge technology talks or marketing campaigns to highlight the capabilities of the new product. If the list of new features do not meet end-users' expectations, the company may face a potential loss in product market share, and a loss in opportunity to implement other features that may be more useful to their users.

This paper discusses the process of how UX researchers may help product management teams with feature selection. In the following sections, we discuss some guiding principles on feature selection research, the challenges faced by the UX researchers, and one of our humble suggestions to tackle the problem. We also present a relevant case study where we successfully applied and adapted the method for a more accessible survey experience for respondents with disabilities.

## GUIDING PRINCIPLES

Since new product development is a multi-million-dollar investment, we suggest the following UX research guiding principles to ensure that research insights are statistically sound and applicable to the user base when informing product development decisions (e.g., Chapman & Rodden, 2023; Goodman & Kuniavsky, 2012). Specifically, we came up with four principles that can guide UX research design.

- First, UX research participants should be representative of the user population of interest. The underlying assumption is that the user base of the new product would not dramatically change from one version to the next. As an example, if we are to conduct a UX research study with Android Tablet users, we will first need to know the proportion of different age ranges, genders, and physical locations of the users. In the participant recruitment, we should strive to assemble a group of participants with the attributes in the same proportion.

- Second, a large enough sample size should be achieved for high stake research questions. Sample size should also be determined prior to the execution of the study to avoid chances of p-hacking or type I statistical errors (Ioannidis, 2005; Button, K., Ioannidis, J., Mokrysz, C. et al, 2013). In general, the sample size can be determined by a power analysis. The calculation depends on the statistical method of choice, the effect size, the variations in the responses, and so on. Representativeness of the sample to the user base should also be considered. An insufficient sample size, regardless of whether the resulting statistical tests are significant, is highly unlikely to be representative of the user base.

- With the backdrop of new feature selection, a meaningful cutoff that limits the number of new features is needed. A useful baseline for comparison is the chance level. In the case of 20 potential new features, there is a 5% chance that UX research participants will pick any options as their top feature. If a feature is to be deemed as having a high potential, it should be chosen by at least 5% of the research participants. This also means that only around 4 to 5 new features will likely cross the bar for further consideration.

- Lastly, when testing the attractiveness of the features, attention should be paid to avoid potential bias. In some cases, the new features may be hard to communicate, and researchers may provide videos or animation explanations prior to the prioritization exercise. In other cases, the more complicated features may lead researchers to use jargon, or more text in the description. These may give some features more attention in the exercise, and hence bias the outcome.

## CHALLENGES

In product feature selection exercises, common surveying techniques include top choice selection, ranking, and rating question types (McCarty & Shrum, 2000). With the goal of understanding the relative importance of feature choices from the perspective of consumers, there are critical differences in the amount of information, level of inference, and pitfalls inherent in these standard techniques.

- Large sample size requirement: Techniques that only require respondents to make a single preference selection from a list of feature contenders requires a large sample size. For a simple two-choice poll, a sample in the order of 1000 participants is usually used to help draw reliable conclusions at the 95% confidence level (Wikipedia, retrieved 2025). In the case of early feature testing, dozens of potential features are usually included in the test. A Bernoulli-based statistical test would require a large number of participants.

- Poor discrimination between alternatives: It can be challenging to infer how important each feature is relative to all candidates on a paired comparison level with single preference selection type questions. On the

- other hand, with rating questionnaires (i.e., value assignment on a likert scale), researchers may face a common measurement problem where there may be insufficient variation in rating responses between options, leading to poor discrimination between alternatives.

- Response bias risks: With rating type questions, respondents tend to show acquiescence bias or fence sitting behavior, giving similarly high or neutral ratings across the board. This can distort data quality and make it difficult to accurately identify preferences.

- Cognitive load: Traditionally, ranking questions provide discrimination between alternatives as respondents are required to give a priority order to featured options; however, performing a ranking of all options can be a difficult task, and the longer the list of options for comparison, the more time it takes to complete (Feather, 1973). Keeping in mind it is not uncommon for product teams to be considering up to a dozen or more features as a research inquiry.

In contrast, the MaxDiff method asks respondents to choose the best and worst options from a small set of varying product features over several trials (Louviere, 1993; Louviere, Flynn, & Marley, 2015). This determines feature preference without overwhelming respondents with ranking a large number of items. This process enables researchers to generate a reliable priority list of potential features using preference share data from a feasible sample size of approximately 300 - 500 participants.

## PROPOSED BEST PRACTICE

As UX researchers, we had numerous research experience with feature selection tasks. Our experience is that a multinomial logistic regression, a method usually marketed as MaxDiff, is reasonably well suited for feature prioritization research questions.

In a MaxDiff survey, respondents are asked the same prompt regarding their preference on featured items multiple times. An example prompt would be: "Among the following potential TV streamer features, which would be the most useful to you?" Each time the question is presented, up to 3-5 features would be shown for respondents to select one that best matches the prompt. The features are a randomly selected subset of the full list which can comprise a dozen features or more in total. Each respondent would go through around 10 rounds of selection.

The task is scalable once the survey is set up. Hundreds of respondents can be invited and recruited to participate in the survey. Once all the respondents are collected, a multiple logistic regression can be applied to the data. There are many options to analyze the data with multinomial logistic regression on the market, including Qualtrics, Sawtooth (Orme, 2009), or open source solutions such as Bayesm (Rossi, Allenby, & McCulloch ,2005).

The outcome of the logistic regression is usually a list of feature preferences. The list establishes a "percentage share" of each feature. In simple terms, a percentage share denotes the percentage of all respondents who selected an option as their top feature, among the full list. The "percentage share" is on a linear scale, so that features with 10% percentage share are twice as popular as those with 5% percentage share. In totality, the percentage shares add up to 100%.

With a large enough sample size of 500 respondents or more, and careful crafting of the feature descriptions in the survey, this method satisfies the guiding principles we discussed earlier.

## CASE STUDY

Beyond its advantages for robustly assessing consumers' preference ranking for a given set of candidate choices, MaxDiff can be formatted to work with assistive tools (i.e., screen readers) to provide an accessible survey experience.

### An Accessible Application of the MaxDiff

The MaxDiff survey approach is typically presented in a matrix format, necessitating respondents to consider two prompts (e.g., best/worst option) for each feature item in the matrix question set. The layout of the matrix format makes it difficult for screen reader users to parse through, adding to the time it takes to consider each matrix line item. This contributes considerable time to the overall survey completion time and can result in respondent fatigue.

An alternative approach that lends itself more accessible to screen reader users is a single prompt presentation style wherein respondents need only select the best (e.g., most important or most preferred) candidate from a subset of choices on each trial. This "Best-Only" MaxDiff provides a simplified form that is effectively a single select, multiple choice questionnaire in experience for respondents.

### Best-Only MaxDiff Case Study

As a case study of the Best-Only approach in application, we present our tablet feature prioritization research that has explored product directions with people with disabilities. We sought to build experiences for tablet users with disabilities, starting with ideation workshops with our team members and participating users to surface accessibility needs that can be best met with tablet solutions. With the breadth of ideas germinated from workshops, our team had worked to refine the ideas to focus on 18 candidate concepts. Note the example concepts discussed here are hypothetical.

Although the potential candidate ideas have been narrowed down, it is still far from what a team can feasibly pursue. It is at this juncture that a MaxDiff can be best applied to prioritize the concepts based on importance from the perspective of tablet users. With the application of the Best-Only method, respondents were instructed to only select the tablet feature they find to be most important, as opposed to selecting their most important and least important feature, thereby reducing the cognitive decisions required on each trial by half (see Fig. 1). We further simplified the survey experience by limiting the number of features presented to three per trial.

Figure 1: **Illustrations of the typical MaxDiff matrix format (top) and the Best-Only MaxDiff single-select answer format (bottom).**

Although there may be features that may hold great importance and improve the product experience for everyone, there may also exist high variation in utility and appeal of any feature relative to the accessibility needs unique to certain cohorts and individuals. In this regard, MaxDiff is a tool best suited to examine the relative preference or importance of a given set of feature choices at an overall and subgroup level. With the research guiding principles we discussed earlier, we recruited the 200 - 350 individuals to represent our user base. These research cohorts included individuals with low vision/blindness, individuals with hearing loss, individuals with mobility impairments, older adults who are 55 or older, and adults who are between 18 and 55 years of age. To account for the differing accessibility needs and interests of various user groups, we used MaxDiff analysis to determine if particular concepts were more important to certain segments of users. By recruiting and analyzing at the subgroup level, we were able to identify tablet features that resonated with particular cohorts (for example, individuals with a visual impairment rank Voice Control higher in importance than sighted individuals) as well as any concepts that have more universal appeal.

## CONCLUSION

This paper discusses how we tackle product feature prioritization, an essential step in the product development cycle that should be informed with UX research. Using the UXR PoV Playbook framework, we described the constraints of existing research methods, chiefly the cognitive strain placed on respondents in both moderated and unmoderated procedures prevalent in the industry. As an alternative, we proposed a methodology that has been tested and applied to support the product development team in learning from reliable consumer input in their feature selections. In the theme of a UXR PoV "Play", we illustrated how the MaxDiff method could be adapted to suit participants with accessibility needs. The adapted MaxDiff took less time for the survey participants to complete, and hence had the potential to improve data quality.

## KEYWORDS

Feature prioritization, MaxDiff, Multinomial logistic regression, User Experience Research, Point of View

UXR Point of View on Product Feature Prioritization					CHI'25, May, 2025, Yokohama, Japan## REFERENCES

[1] Beyer, H., & Holtzblatt, K. (1990). Contextual design: Designing customer-centered systems. Morgan Kaufmann Publishers.

[2] Button, K., Ioannidis, J., Mokrysz, C. et al. (2013). Power failure: why small sample size undermines the reliability of neuroscience. Nat Rev Neurosci 14, 365–376. https://doi.org/10.1038/nrn3475

[3] Chapman, C., & Rodden, K. (2023). Quantitative User Experience Research: Informing Product Decisions by Understanding Users at Scale. Apress.

[4] Dogan H., Giff S., & Barsoum R. (2024). User Experience Research: Point of View Playbook. In Extended Abstracts of the CHI Conference on Human Factors in Computing Systems (CHI EA '24). Association for Computing Machinery, New York, NY, USA, Article 537, 1–7. https://doi.org/10.1145/3613905.3637136

[5] Feather, N. T. (1973). The measurement of values: Effects of different assessment procedures. Australian Journal of Psychology, 25, 221-231.

[6] Goodman E., & Kuniavsky, M. (2012). Observing the User Experience: A Practitioner's Guide to User Research. Morgan Kaufmann.

[7] Ioannidis, J. P. (2005). Why most published research findings are false. PLoS medicine, 2(8), e124.

[8] Louviere, J. J. (1993), The Best-Worst or Maximum Difference Measurement Model: Applications to Behavioral Research in Marketing. The American Marketing Association's 1993 Behavioral Research Conference, Phoenix, Arizona.

[9] Louviere, J. J., Flynn, T. N., & Marley, A. A. J. (2015). Best-Worst Scaling: Theory, Methods and Applications. Cambridge University Press.

[10] McCarty, J. A., & Shrum, L. J. (2000). The measurement of personal values in survey research: a test of alternative rating procedures. The Public Opinion Quarterly, 64, 271-298.

[11] Orme, Bryan (2009), Using Calibration Questions to Obtain Absolute Scaling in MaxDiff. SKIM/Sawtooth Software European Conference, Prague, Czech Republic.

[12] Remy, L., Lau, J., Brumer, M., & Lu, M. (2023). What's best for users with a11y needs is best for all: Accessible stack ranks with best-only Max Diff [SurveyCon, Google Internal Conference].

[13] Rossi, P. E., Allenby, G. M., & McCulloch, R. (2005). Bayesian statistics and marketing. John Wiley & Sons.

[14] Simon, H. A. (1982). Models of bounded rationality (Vols. 1-2). MIT Press.

[15] Tetlock, P. E., & Gardner, D. (2015). Superforecasting: The art and science of prediction. Crown.

[16] Wikipedia (2025). https://en.wikipedia.org/wiki/Sample_size_determination#Estimation. Retrieved February 24, 2025.